# Substantial Local Variation of Seebeck Coefficient in Gold Nanowires

Pavlo Zolotavin[1], Charlotte Evans[1], Douglas Natelson[*,1,2,3]

[1]Department of Physics and Astronomy, Rice University, 6100 Main St., Houston, Texas 77005, United States

[2]Department of Electrical and Computer Engineering, Rice University, 6100 Main St., Houston, Texas 77005, United States

[3]Department of Materials Science and NanoEngineering, Rice University, 6100 Main St., Houston, Texas 77005, United States

[*]E-mail: natelson@rice.edu.

**Abstract**

Nanoscale structuring holds promise to improve thermoelectric properties of materials for energy conversion and photodetection. We report a study of the spatial distribution of the photothermoelectric voltage in thin-film nanowire devices fabricated from single metal. A focused laser beam is used to locally heat the metal nanostructure via a combination of direct absorption and excitation of a plasmon resonance in Au devices. As seen previously, in nanowires shorter than the spot size of the laser, we observe a thermoelectric voltage distribution that is consistent with the local Seebeck coefficient being spatially dependent on the width of the nanostructure. In longer structures, we observe extreme variability of the net thermoelectric voltage as the laser spot is scanned along the length of the nanowire. The sign and magnitude of the thermoelectric voltage is sensitive to the structural defects, metal grain structure, and surface passivation of the nanowire. This finding opens the possibility of improved local control of the thermoelectric properties at the nanoscale.

**Introduction**

Control of material structure and dimensions on the nanoscale offers opportunities for engineering and enhancement of thermoelectric properties.[1–4] Realizing this potential requires a detailed understanding of the variation of thermoelectric properties at the nanoscale and its underlying mechanisms. One approach to enhance the thermoelectric figure of merit typically employed in semiconductor nanowires is to reduce thermal conductance by increasing phonon boundary scattering while the electrical conductivity remains minimally affected.[5–8] This approach is less beneficial in thin metal films and nanowires because scattering from surfaces reduces the electron and phonon mean free path, which in turn lowers electronic and "phonon drag" contributions to the Seebeck coefficient, $S$.[9–14] Differences between the local values of $S$ in nanowires with varying width were recently used to develop microscale thermocouples fabricated from single metal.[15,16]

Given a nanoscale origin of the thermoelectric response in single metal thermocouple devices, a more localized approach is necessary to study these nanostructures. In most experiments, the local temperature increase is created by a separately controlled resistive heater, which leads to an averaged response resulting in the loss of information about the local values of Seebeck coefficient. Spatial variation of the thermoelectric properties can also be studied using scanning laser microscopy, with the focused beam as a localized heat source.[17,18] This method has been applied to examine position-sensitive photothermoelectric (PTE) properties in suspended carbon nanotubes,[19,20] thermo-voltages in mechanical break junctions,[21] and novel photodetectors.[22–27] This technique is of great interest to study PTE response in plasmonic metal nanostructures, because plasmon excitation can create highly localized, non-thermal electronic distributions that can be potentially harnessed to improve thermoelectric response.[28,29]

We report a detailed study of the PTE voltage generated in thin-film gold nanowires at room temperature and at substrate temperatures down to 5 K. A focused laser spot produces a localized heat source due to the direct light absorption and excitation of the transverse plasmon resonance of the nanowire. As the laser is scanned across the device, the thermally generated open circuit photovoltage is recorded as a function of beam position. A typical magnitude of the PTE voltage is in the range of 1-5 µV per mW of laser power on the sample. Local temperature increase in the device at the maximum available laser power was previously quantified using a bolometric approach to range from ~10 K at room temperature to ~140 K at low temperature.[30,31] As reported recently,[29] We observe the PTE voltage generated in nanostructures with nanowires shorter than the laser spot size to be qualitatively similar to the behavior observed in single metal thermocouples.[15,16] For longer nanowires, we find extreme spatial variability of the PTE voltage. We observe multiple sign changes, large sensitivity to structural defects, grain boundaries, and surface conditions – all are unexpected for an ordinary thin-film metal nanowire with relatively low sheet resistance. We attribute the spatial variability of the PTE voltage to the changes in the local value of the Seebeck coefficient of the nanowire caused by the specifics of the grain structure and surface passivation. Understanding these findings may lead to new opportunities for the possible modification and improvement of the local thermoelectric properties of metals at the nanoscale.

**Results and discussion**

A typical device studied in this work consists of the nanowire contacted by the larger 10 µm wide electrodes fabricated from the same material, Fig. 1a. We begin with a discussion of such a short nanowire as a point of comparison for our main results obtained on longer nanowires. The device geometry and experimental measurement approach is similar to the devices without nanogaps described in previous publications.[30–32] A focused laser beam of the scanning microscope produces the local temperature increase necessary to generate an open circuit thermoelectric voltage across the nanostructure. The thermoelectric voltage difference between the ends of the wire that is heated somewhere in the middle, while the ends are kept at identical temperatures, is given by

$V = \int_{-l}^{l} S(x,T)\nabla T(x)dx$, where $S$ is the location and temperature-dependent Seebeck coefficient of the material and $T$ is the local temperature. On the nanoscale, $S$ can be locally modified by the changes in the electron mean free path.[15,33–36] In this case, the Seebeck coefficient of thin metal film is width-dependent. As we reported recently,[29] the change in the width of the device at the ends of the nanowire creates a small difference between the Seebeck coefficients of the nanowire and the fan-out electrode. This junction acts as an effective thermocouple temperature sensor when the beam is positioned in this spot, producing a maximum or minimum in the PTE voltage map, Fig. 1d When the laser beam is positioned in the middle of the nanowire, both the temperature profile and the geometry of the nanostructure are symmetric and therefore no PTE voltage is observed (Fig. 1c,d). The absolute value of the PTE voltage has linear laser power dependence at room temperature, Fig. S2. In devices fabricated from gold, when the device is illuminated with the light polarized perpendicular to the long dimension of the nanowire, additional heating is present due to the excitation of the plasmon resonance. Using longitudinally polarized light (polarization angle is denoted as 0°) one can eliminate the plasmonic contribution to heating, with the local temperature increase reduced to that produced by direct optical absorption and so the observed voltage should decrease, as demonstrated in Fig. 1d.

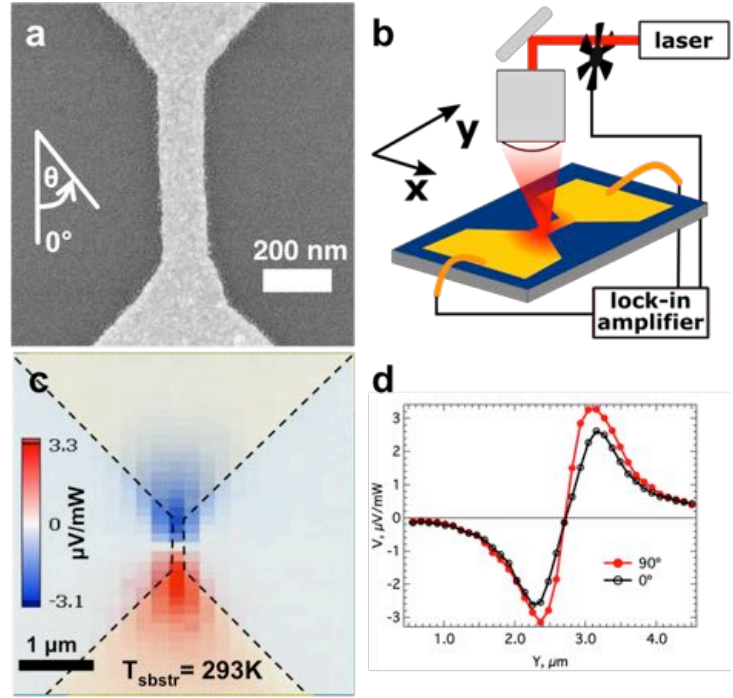

**Fig. 1**. Photothermoelectric voltage maps in Au/Ti devices with nanowire length comparable to the focal spot size, similar to that reported recently.[29] (a) Scanning electron microscope (SEM) image of a typical bowtie device. The width of the nanowire 135 nm, which allows for the excitation of the transverse plasmon resonance with 785 nm laser polarized perpendicular to the nanowire. (b) Schematics of the experimental setup. (c) Spatial distribution of the PTE voltage in Au/Ti device at room temperature. The map is overlaid with the false colored SEM image of the device. (d) Voltage along the vertical centerline of the device from (c) for the polarization perpendicular (filled circles, 90°) and parallel to the nanowire (empty circles, 0°). For panels (c) and (d) voltage is reported in units of µV per mW of laser power on the sample. The bottom electrode is connected to the -B input of the voltage amplifier.

Building on previous modeling of the local temperature rise, $ΔT$, that was previously inferred using a bolometric technique,[30,31] we set up a simplified 2D model in COMSOL Multiphysics to estimate the value of the thermoelectric voltage (see Electronic Supplementary Information (ESI) for details) from a known thermal gradient across the device. The magnitude and spatial dependence of the observed thermoelectric voltage is consistent with the proposed mechanism of width dependent Seebeck coefficient. A roughly estimated local $\Delta S$ of ~4% and $\Delta T$~15 K is sufficient to generate ~1 µV voltages at the ends of the 10 µm long nanowire under 10 mW of simulated illumination power, which corresponds to ~0.1 µV/mW. In the short nanowire experiment shown in Fig. 1c, d, the typical measured PTE voltage is around 1 µV/mW, implying that actual $\Delta S$ due to width-dependent boundary scattering should be larger than 4%. In addition to the room temperature data, we also performed experiments at substrate temperature of 5 K, Fig. S3. At low temperature the absolute value of Seebeck coefficient is reduced and therefore the value of $\Delta S$ is also expected to be smaller. This reduction in $\Delta S$ is, however, offset by a significantly larger $\Delta T$~140 K at similar laser power level, producing PTE voltages that are comparable in magnitude despite large differences in local and substrate temperatures. As the exact dependence of $S(T,w)$ for the thin metal film is not known, it is difficult to develop a more comprehensive computational model to precisely reproduce the experimental findings. A good agreement with a qualitative model, nevertheless, corroborates the thermoelectric origin of the observed photovoltage.

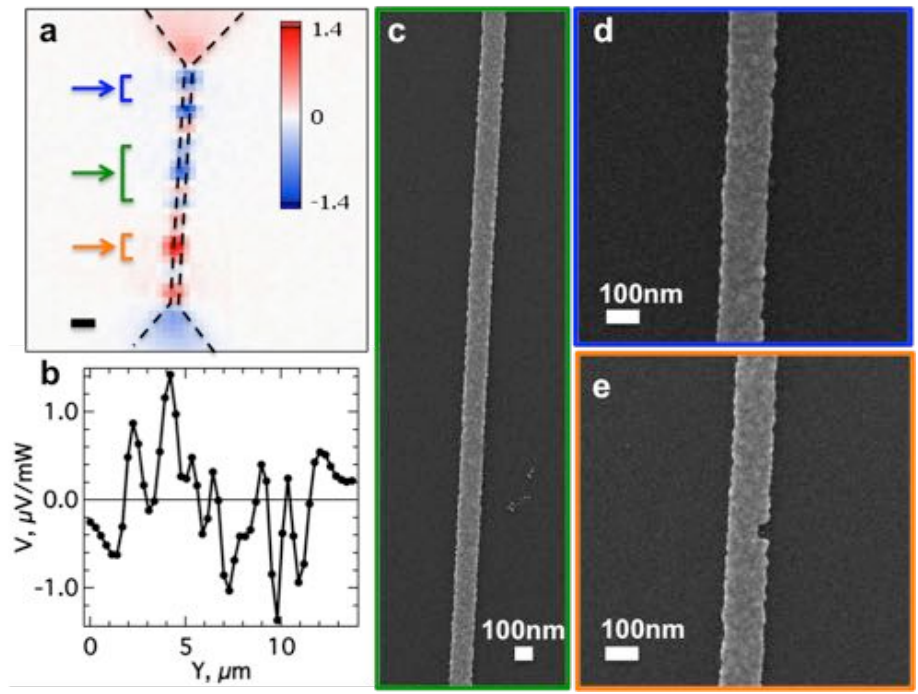

**Fig. 2**. Bowtie devices with long nanowires display extreme spatial variability of PTE voltage along the nanowire. (a) PTE voltage map of a typical 10 µm long and 100 nm wide Au/Ti device, in units of µV per mW of laser power on the sample. Scale bar is 1 µm. Substrate temperature is 5 K. (b) Variation of the PTE voltage along the length of the device. (c) SEM image of the central part of the nanowire, displayed area is highlighted in (a) by the arrow in the center. (d),(e) the same as (c), but in the top and bottom section of the device.

The length of the nanowire in Fig. 1 is 0.7 µm, which is smaller than the laser beam spot diameter of 1.8 µm. For this geometry, the resulting PTE voltage map is strongly influenced by the location of the fan-out electrodes. To investigate the spatial dependence of the PTE voltage along the nanowire itself, we fabricated 10 µm long devices, Fig. 2a. The magnitude and sign of the PTE voltage consistently showed strong spatial dependence, switching sign multiple times along the length the devices, Fig. 2b. The nanowire in Fig. 2 is narrower than optimal width for plasmon heating, which explains smaller PTE voltages than that of a short device. The shape of the PTE voltage distribution along the nanowire has weak dependence on the substrate temperature and laser power (Fig. S4), remaining essentially unchanged in the substrate temperature range of 5 K to 165 K and $\Delta T$ from 35 to 145 K. Spatial variation is observed for devices at room temperature, Fig. S5, and for wider devices in which heating occurs only by direct laser absorption excluding plasmon resonance excitation, Fig. S6. Some of the features in the PTE map are correlated with the position of large structural defects in the nanowire as demonstrated in Fig. 2e. The local variation of width acts as the single metal thermocouple described earlier, but now on a much smaller length scale. As a result, the PTE voltage map develops a characteristic min/max feature close to the location of the defects in addition to the min/max pattern created by the thermocouples at the ends of the nanowire itself, Fig. S5. The surprising observation, however, is that the sections of the nanowire without clearly visible defects also demonstrate PTE voltage sign variation, Fig. 2c,d and Fig. S7. This result suggests that there are other sources that can affect the local value of Seebeck coefficient along the metal nanowire.

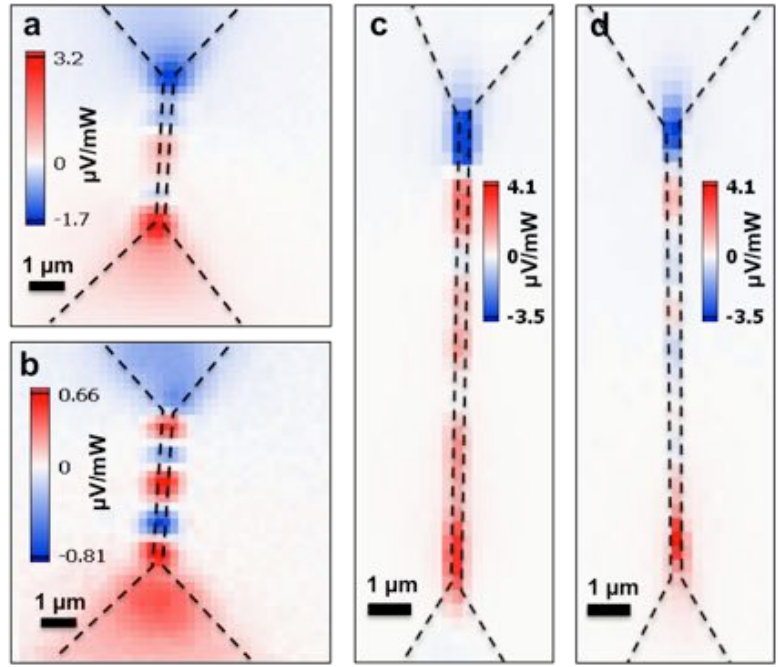

**Fig. 3**. Sensitivity of the PTE voltage maps to the local grain structure probed by annealing. (a) Initial PTE voltage map measured at room temperature. (b) The same device annealed by passing current before the onset of electromigration. We note that the magnitude of the PTE voltage is significantly reduced after annealing for this experiment. The resistance of the device decreases by ~2-3% after current annealing. (c) PTE voltage map for a 10 µm Au/Ti device. (d) The same device after annealing at 200 °C for 3 hours in Ar atmosphere. All data acquired at room temperature.

Mild annealing of the nanowire by passing electrical current that is large enough to slightly reduce the total device resistance, but insufficient to start the electromigration process, can substantially change the PTE voltage map. Fig. 3a,b demonstrates the PTE voltage map before and after such annealing. The PTE voltage sign distribution is altered, while inspection of the SEM images before and after annealing does not show visible changes in the grain structure. The magnitude of the current for this type of annealing depends on the substrate temperature and varies slightly between devices, because the onset of electromigration (when current clearly affects material structure) strongly depends on the local grain structure of the nanowire. Annealing provides the same overall effect for different nanowires, lowering the total resistance without changing the geometry of the device in any way noticeable from SEM images. The PTE voltage map is also affected by conventional annealing at high temperature, Fig. 3c,d. In this case the $\Delta S$ variation in the nanowire region initially dominates the initial PTE voltage map, but after annealing the contributions from the ends of the nanowire are more prominent. Annealing changes the position of defects and grain boundaries in the crystal structure of the nanowire, which indicates that the local value of Seebeck coefficient is depended on the location and specifics of the grain size distribution and their boundaries.

The electron mean free path in thin metal films is reduced from the bulk value by surface scattering and is comparable to the thickness of the film. Using the Drude model, we estimate the magnitude of the electron mean free path to be in the 12 to 20 nm range depending on the substrate temperature, Fig. S8. As the electron mean free path is smaller than the bulk value (39 nm) and is comparable to the nanowire thickness, the surface scattering plays important role in the diffusive transport of charge carriers in the nanostructure. The specific value of the Seebeck coefficient is determined by the details of the charge carrier transport through the material,[37] and, therefore, one can expect some dependence of PTE voltage map on surface conditions. The importance of the surface is revealed in experiments in which the device is covered with self-assembled monolayer (SAM) of benzyl mercaptan. In this case, current annealing does not change the PTE voltage map, Fig. 4a,b. Once the SAM is removed, however, the map is altered, Fig. 4c. Additionally, $O_2$ plasma treatment of the nanowire strongly effects the PTE voltage distribution, Fig. S9. In this situation, scanning with large laser intensity produces a PTE voltage map that is considerably different from a typically observed pattern. A subsequent scan then changes the distribution similar to that obtained after high temperature annealing. The effects of surface passivation and $O_2$ plasma treatment are observed both at room temperature and substrate temperature of 5 K. These results demonstrate that the Seebeck coefficient of thin metal nanowires is sensitive to the surface passivation. Determination of the exact physical origin of this effect will require experiments with systematic surface treatment designed to shift the work function,[38] and will be a subject of future publications.

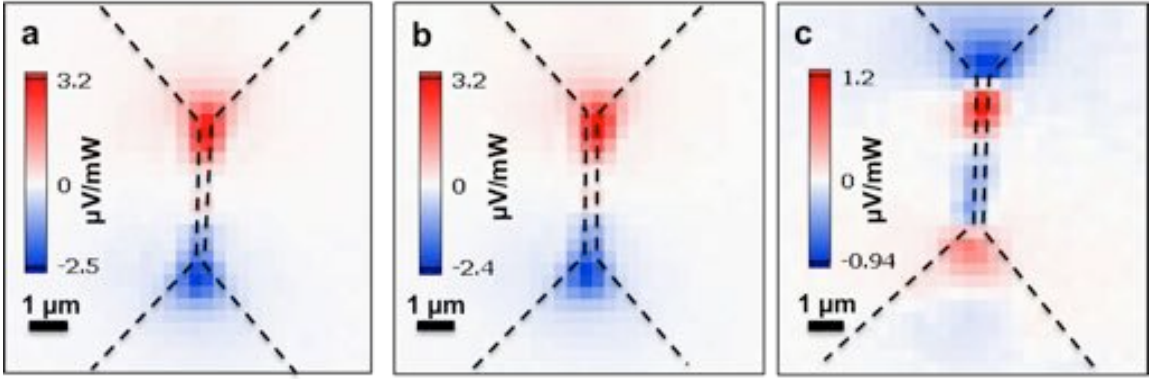

**Fig. 4**. Effect of the surface passivation on the PTE voltage map is significant. (a) PTE voltage map for a device covered with benzyl mercaptan SAM acquired at substrate temperature of 5 K. (b) The same device after current annealing does not display changes in the spatial distribution of PTE voltage. (c) PTE voltage map for the device from (b) after SAM is removed using dilute solution of $NaBH_4$.

Substantial variation of the local value of Seebeck coefficient in long nanowires is not specific to devices fabricated from gold. We observe similar effects in devices made from Ni, AuPd alloy and Au devices without Ti adhesion layer, Fig. S10, S11. When comparing PTE maps for nanowires fabricated from different materials, we see that magnitude of the PTE voltage generally follows the Seebeck coefficient of the metal used in the experiment. Au devices without Ti adhesion layer have the smallest PTE voltage of ~2-5 µV/mW and the smallest value of absolute Seebeck coefficient for bulk Au, 1.5 µV/K. For bulk Ni the Seebeck coefficient is -20 µV/K, which corresponds to the PTE voltage of ~50 µV/mW, Fig. S11a. Close to room temperature, the 60/40 AuPd alloy has the largest Seebeck coefficient -35 µV/K,[39] and the largest PTE signal ~100 µV/mW. We note that AuPd devices fabricated with Ti adhesion layer had anomalously small values of the PTE voltage, Fig. S11b, which is most likely the result of the alloying of AuPd with Ti metal.

Spatial resolution of the experimental setup is determined by the convolution of the size of the laser spot, defect size, and the pixel density during the PTE voltage map acquisition. We estimate the effective spatial resolution of 20 nm structural defects (as determined within limitations of the post-measurement scanning electron microscopy) in the Seebeck response at 0.3 µm. This parameter could be improved by super resolution microscopy methods[40] in attempts to correlate the PTE voltage map with the detailed map of the grain structure of the nanowire, provided the latter is known. A better technique to study the correlation between local crystal structure and thermoelectric properties would be to fabricate a single-crystal nanowire device with a single grain boundary defect in the known location. This approach will circumvent the limitations of low spatial resolution. These experiments will be the subject for future studies.

We now consider candidate mechanisms responsible for the observed spatially varying PTE voltage. As mentioned previously, PTE voltages in single metal nanostructures can be explained by the spatial variations of Seebeck coefficient that arise due to the changes of the electron mean free path between sections of thin metal films of different width or thickness.[15,16] This effect has been used to develop IR photodetectors and two-dimensional arrays for local temperature measurement.[34–36] Unlike these earlier works, which focused on the large-scale illumination or uniform lithographically defined heaters, the present localized laser scanning microscopy reveals large spatial variability of the PTE voltage in long metal nanowires without noticeable width variation or other changes in structure that could cause strong local variation in mean free path.

In addition to the grain boundary structure and surface passivation causing local variations of $S$, an alternative source of extreme spatial variability of the PTE voltage in long devices could be local variation of the local value of absorption coefficient (and hence the heated temperature distribution) of the metal nanostructure. After a conservative quantitative analysis (see ESI for details), we dismiss this possibility; the variation in material properties necessary to produce a reversal in the temperature gradient would be unphysical.

The specific positions of impurity atoms can act as another source of spatial variability of thermoelectric properties. The thermopower of gold is known to be sensitive to impurities and can even change sign at low temperatures in the presence of a few ppm of Fe.[37,41] Alloying with a Ti adhesion layer could also potentially modify the local value of the Seebeck coefficient. A comparison of the PTE voltage maps at different laser power and at substrate temperatures, Fig S4, reveals no change in the PTE voltage distribution map, implying that the Seebeck coefficient remains of the same sign for $T > 40$ K. The similar spatial variability in PTE voltage in AuPd and Ni nanowires further eliminates impurities as the main mechanism behind the observed variability of PTE voltage.

We suggest as a candidate explanation grain-to-grain variability in $S$ due to crystallographic orientation and states at grain boundaries. The value of Seebeck coefficient depends upon the details of how the charge carriers scatter during the diffusive transport through the material.[37] In metals, the dominant contribution to the electronic Seebeck coefficient is proportional to the energy derivative of the electrical conductivity.[37] It is therefore likely that changes in the local electronic properties would define the magnitude of the variation $\Delta S$. For example, scanning tunneling microscope investigation of surface thermoelectric properties of gold revealed large variation in local thermoelectric voltage while crossing the surface terrace steps.[42–45] Differences between crystallographic orientation of individual grains of the nanowire can create abrupt changes in the Fermi energy surface while crossing the grain boundaries, even though the overall variation of work function for gold is modest for different crystallographic orientations.[46] The electron mean free path in thin metal films is strongly constrained by the film thickness and therefore the details of surface scattering will also affect the thermoelectric properties of thin metal films.[47–49] We have taken the model developed previously to infer the temperature profile of $\Delta T$ along the short nanowire,[30] and we have updated it to account for the specific longer nanowire geometry of the present experiments. We then estimate the magnitude of $dT/dx$ to be as large 10 K/μm at 10mW of incident laser radiation. This would imply that, to produce the PTE maps observed in the experiments, the variation in the local value of the Seebeck coefficient would have to be in the 10-20% range, Fig. S12. We note that the change the PTE voltage sign around the defect in spatial maps does not imply a change in the sign of the Seebeck coefficient. A local variation in S creates a "thermocouple" pattern analogous to the one observed in Fig. 1. As the sign of the ΔS is fixed by material properties and the electrical connection to the device remains unchanged during the data acquisition, the different sign of the PTE voltage will be observed depending on which side of the local variation ΔS is heated.

The model developed for the calculation of the local temperature does not take into account size effect and possible violation of the Fourier transport laws for low temperature and devices with nanoscale dimensions.[50–52] The model however agrees well with bolometric measurements of changes in device conductance, and also with previous independent estimates from the ratio of the anti-Stokes to Stokes Raman spectra intensity. The inference from the PTE voltage maps of the spatially varying Seebeck response is in fact independent of the details of the temperature model. As the variation in the Seebeck coefficient occurs on the length scale comparable to or smaller than the electron mean free path, a more detailed model that would take into account the nature of electron transport at the single grain scale is necessary to draw more quantitative conclusions.

**Conclusions**

We demonstrate substantial spatial variability in the local thermoelectric properties of thin gold nanowires. This behavior is attributed to the combination of the variations in the local width of the device, grain structure, and surface scattering – contributions to the total value of Seebeck coefficient that were not revealed in the previous spatially averaged experiments. Control of the surface structure and composition offers a new tool for controlling the thermoelectric response at the nanoscale. While the absolute magnitude of the Seebeck coefficient in metals is small for energy harvesting applications, the approach presented in this work could be extended to study the local variation of Seebeck coefficient in semiconductor nanostructures specifically designed for high thermoelectric efficiency.[53–55] This work is also important for enhancing the performance of photodetectors based on thermoelectric transduction and improving our understanding of the role of local material properties on thermoelectric response of metal nanowires.

**Experimental details**

All devices were fabricated on n-type Si wafers with a 200 nm of thermally grown oxide layer. Prior to the e-beam lithography for nanowire device fabrication, a set of Au/Ti contact pads for wire bonding was evaporated on the substrates using shadow mask. Metallization layers were deposited using e-beam evaporator. Thickness of the Au nanowire and connecting electrodes was 14 nm with additional 1 nm Ti adhesion layer. For control experiments we also studied Au devices fabricated without the Ti adhesion layer, devices using AuPd alloy, and devices with Ni instead of Au. Each chip contained 24 devices with shared ground. The total number of devices investigated during this project is 194. The results selected for demonstration represent a typical sample behavior. A home-built scanning laser microscope with ability to record spatially resolved Raman spectra maps was used to perform the experiments.[30,56] The samples were kept in high vacuum of the closed-cycle optical cryostat (Montana Instruments). A mechanical chopper at frequency of 287 Hz was used to modulate laser light for lock-in amplifier voltage measurement. Unless specially noted, the PTE voltage distributions are recorded with laser polarization perpendicular to the long dimensions of the nanowire (angle assignment of 90°) to provide maximum heating. Open circuit voltage was measured using SR560 voltage amplifier. Experiments were conducted with 10mW laser power recorded at the sample, unless specially noted. As an additional crosscheck we measured maps of the closed circuit photocurrent using SR570 current amplifier. The amplitude and sign of the photocurrent were consistent with the voltage maps acquired previously. Devices with SAM were prepared by soaking oxygen plasma pre-cleaned devices in 1 mM solution of benzyl mercaptan in toluene for several hours. The SAM was removed using 0.5M solution of $NaBH_4$ in 50% ethanol.[57]

***Electronic Supplementary Information (ESI) available:*** Results of the modeling, additional experimental data at different substrate temperature, different device geometries, and for devices fabricated from Ni and AuPd alloy.

***Acknowledgments***. P.Z. and D.N. acknowledge support from ARO award W911-NF-13-0476 and from the Robert A. Welch Foundation Grant C-1636. C.E. acknowledges support from NSF GRFP DGE-1450681.

*Electronic Supplementary Information*

# Substantial Local Variation of Seebeck Coefficient in Gold Nanowires

Pavlo Zolotavin[1], Charlotte Evans[1], Douglas Natelson[*,1,2,3]

[1]Department of Physics and Astronomy, Rice University, 6100 Main St., Houston, Texas 77005, United States

[2]Department of Electrical and Computer Engineering, Rice University, 6100 Main St., Houston, Texas 77005, United States

[3]Department of Materials Science and NanoEngineering, Rice University, 6100 Main St., Houston, Texas 77005, United States

[*]E-mail: natelson@rice.edu.

# Estimate of the total voltage using a simplified 2D model of heating in the bowtie nanostructures

A spatial distribution of the PTE voltage across the device could be reproduced using a simplified 2D heat dissipation model that was implemented using a COMSOL Multiphysics. The model described here was adapted from the Supplementary Material in Ref. 1. Heat dissipation was modeled using a 2D geometry that also includes an out-of-plane heat transfer to the $SiO_2$ substrate. The substrate temperature, and left/right boundaries were held at fixed temperature, 293 K. The localized heat source had a Gaussian distribution to simulate heating from the focused (Gaussian spot) laser beam. Additional heating from the plasmon resonance excitation was modeled by including a width dependent heater modulation with the maximum contribution of additional factor of 4 in the nanowire geometry segment. The intensity of the heat source was adjusted to produce a local temperature increase comparable to that inferred experimentally, $\Delta T \sim 10$ K, Figure S1b. Simultaneously with the heat dissipation, the electric potential distribution was calculated. The right boundary of the device was grounded and the rest electrically isolated. The position of the heat source was moved along the centerline of the device to reproduce scanning of the laser beam and the voltage at the left boundary was recorded, Figure S1c. Within the Fuchs-Sondheimer electronic specular reflection model of the resistivity in thin films,[2,3] modified to accommodate reflection from the side walls, the Seebeck coefficient of the film could be written as[4–8]

$$S_f = S_g \left[1 - \frac{3}{8}\frac{(1-p)\lambda_0}{d}\frac{U_g}{1+U_g}\right]$$

where $d$ is the effective film thickness defined as $\frac{1}{d} = \frac{1}{w} + \frac{1}{t}$, $w$ is the width of the nanostructure and $t$ is the thickness of the film, $\lambda_0$ is the electron mean free path in gold, $p$ is the scattering coefficient, $U_g = \left.\frac{\partial \ln \lambda_0}{\partial \ln E}\right|_{E_F}$ and $S_g(T)$ is the Seebeck coefficient of the infinitely thick film approximated here by the bulk value.[9] We estimate the scattering coefficient $p$ as 0.1 by comparison of the resistivity and temperature coefficient of resistance with the previous data for thin films.[7] The value of $U_g$ for thin gold films is close to -0.6.[7] The above equation for $S_f$ is derived in the limit of $d \gg \lambda_0$, however in our case $d \sim 0.4\lambda_0$ and the prefactor 3/8 should therefore be reduced to 0.22. In this model the spatial dependence of the Seebeck coefficient is determined by the dependence of the electron mean free path on the width of the nanostructure. The model produces a spatial distribution of thermoelectric voltage that is qualitatively consistent with the one that would be expected in short devices or long devices without additional spatial variation of $S$ (Figure 1d of the main text). The location of the absolute maximum (minimum) of the thermoelectric voltage is determined by the position of the largest $\Delta S$. For the $S(w,T)$ model presented here it is located close to the nanowire ends as $dS \sim \frac{dw}{w^2}$. It is important to point out that the $\Delta S$ in this model for these reasonable parameters is only ~5% of the bulk value of Seebeck coefficient. An extension of this calculational model to consider local variations in the Seebeck coefficient in longer nanowires is described at the end of the ESI.

In nanowires longer than the size of the focused laser beam, we observe a large variation of the PTE voltage as the laser beam is scanned along the nanowire. We attribute this variability to the changes in the local value of Seebeck coefficient. Alternatively, the experimental results could be explained by the changes in the local value of the temperature gradient due to the differences in absorption coefficient between neighboring sections of the nanowire or the differences in thermal conductance. We estimate that the change in absorption coefficient would require a change in absorption comparable to the magnitude of the absorption coefficient itself. Consider two adjacent patches of the nanowire 100 nm long and 150 nm wide. The incident intensity of 400 kW/cm² (maximum intensity used in the experiment) corresponds to the total incident power on each patch of ~60 µW. Using the total absorption of ~5 %, the total power dissipated in each section is ~3 µW. On the other hand, to drive a 5 K temperature gradient over 300 nm distance at room temperature would require a $\frac{5\,K}{3\cdot 10^{-7} m} \cdot 314 \frac{W}{m\cdot K} \cdot (100\,nm \times 15\,nm) = $ 7.5 µW of thermal power to flow along the wire. This estimate implies that the magnitude of the local difference in absorption large enough to drive a reversal in the sign of the temperature gradient is unphysical. We also argue that it is very unlikely that the local changes in thermal conductance can produce the observed effects. The differences in thermal conductance can affect the local value of the temperature as the heat flows away from the heated section of the nanowire, but not the sign of the temperature gradient, which determines the sign of the observed PTE voltage. However, for the sign of the local temperature gradient (which determines the sign of the observed PTE voltage, in the limit of fixed S) to depend sensitively on spot position would require complex spatial variation of the thermal path.  Given the large electronic component of thermal conductivity in the metal and the fact that the metal films have low sheet resistance (and therefore good electronic homogeneity), this seems unlikely. Large departures from standard thermal transport at the nanoscale are possible,[10] but at present there is no evidence for such effects in these structures under large-spot-size, steady-state illumination.

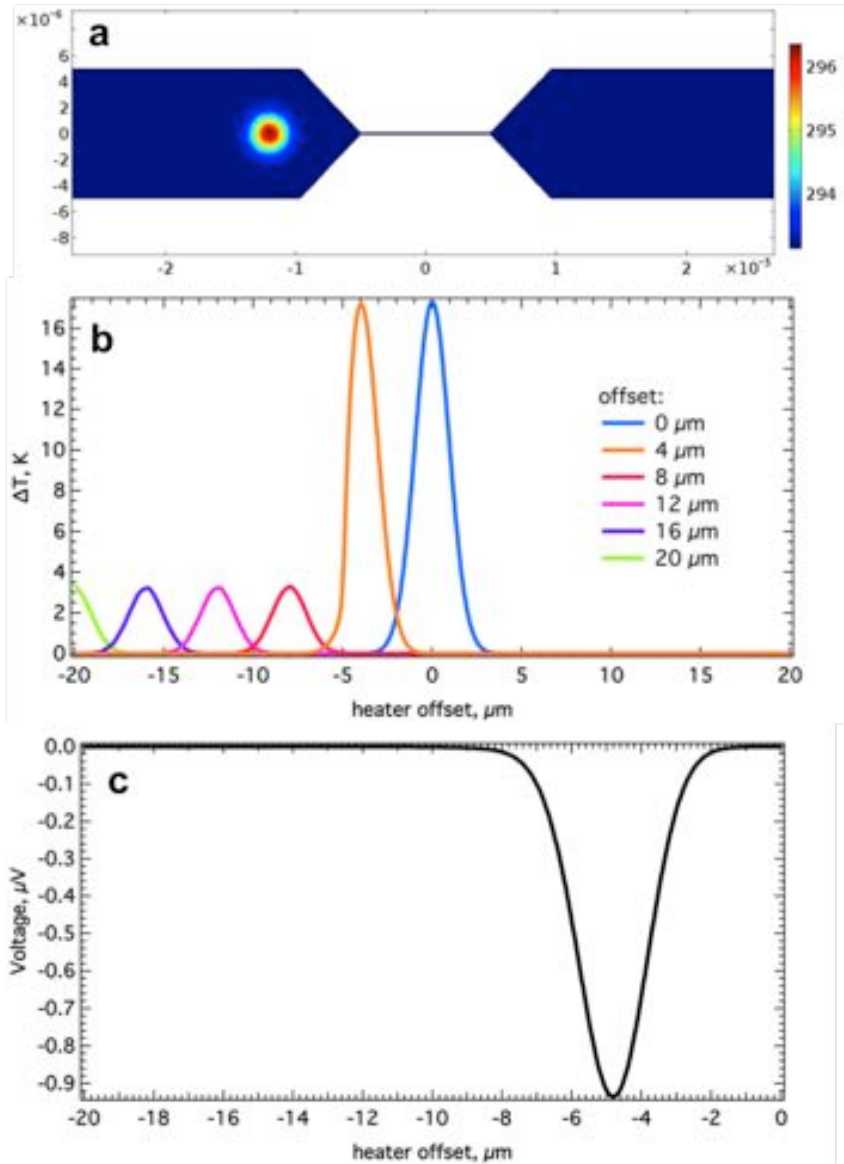

**Fig. S1** Qualitative estimate of the total voltage measured in the experiment using a simplified 2D heating model with an artificial heater. (a) Surface temperature map of the temperature increase for the heater offset by -12 μm from the center of the device. The heater power was adjusted to produce ΔT ~ 15 K in the center of the device. The heater has a modified Gaussian spatial distribution to imitate localized heating from the focused laser beam and plasmon excitation. (b) Temperature profiles along the centerline of the device for different heater offsets. (c) Thermally generated voltage across the device as the function of the heater offset from the center of the device calculated using a spatially dependent Seebeck coefficient.

# Additional experimental data on PTE voltage distribution in gold nanowires at different substrate temperatures

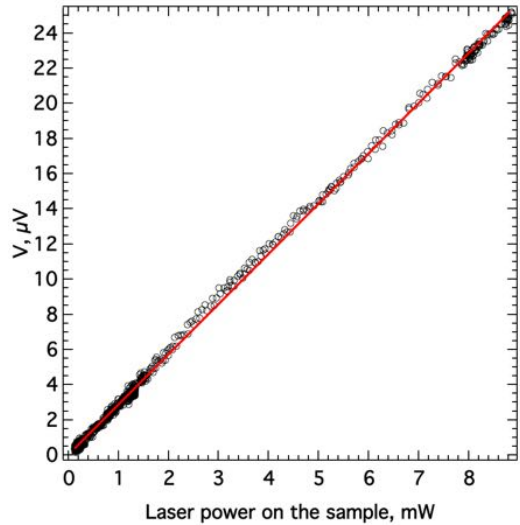

**Fig. S2**. Linear dependence of PTE voltage on laser power recorded at the maximum point of the PTE voltage map acquired at room temperature. Data corresponds to the PTE voltage from Figure 1c.

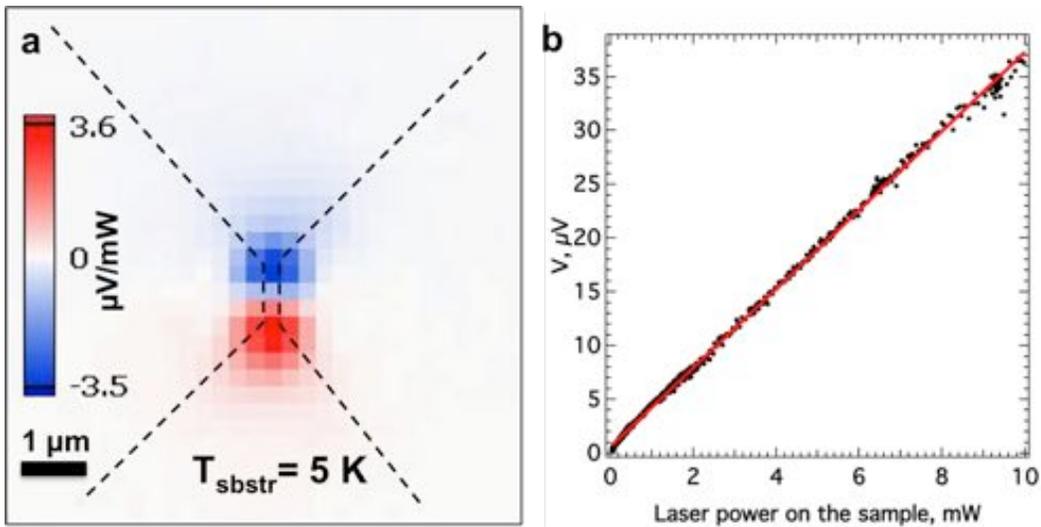

**Fig. S3**. (a) PTE voltage map acquired at substrate temperature of 5 K is similar to the room temperature data. (b) Laser power dependence of PTE voltage deviates from linearity at low temperatures. The non-linearity is a result of the combination of the non-linear laser intensity dependence of temperature increase of the nanostructure at low temperatures and non-linear temperature dependence of Seebeck coefficient in the 5 K to 150 K.

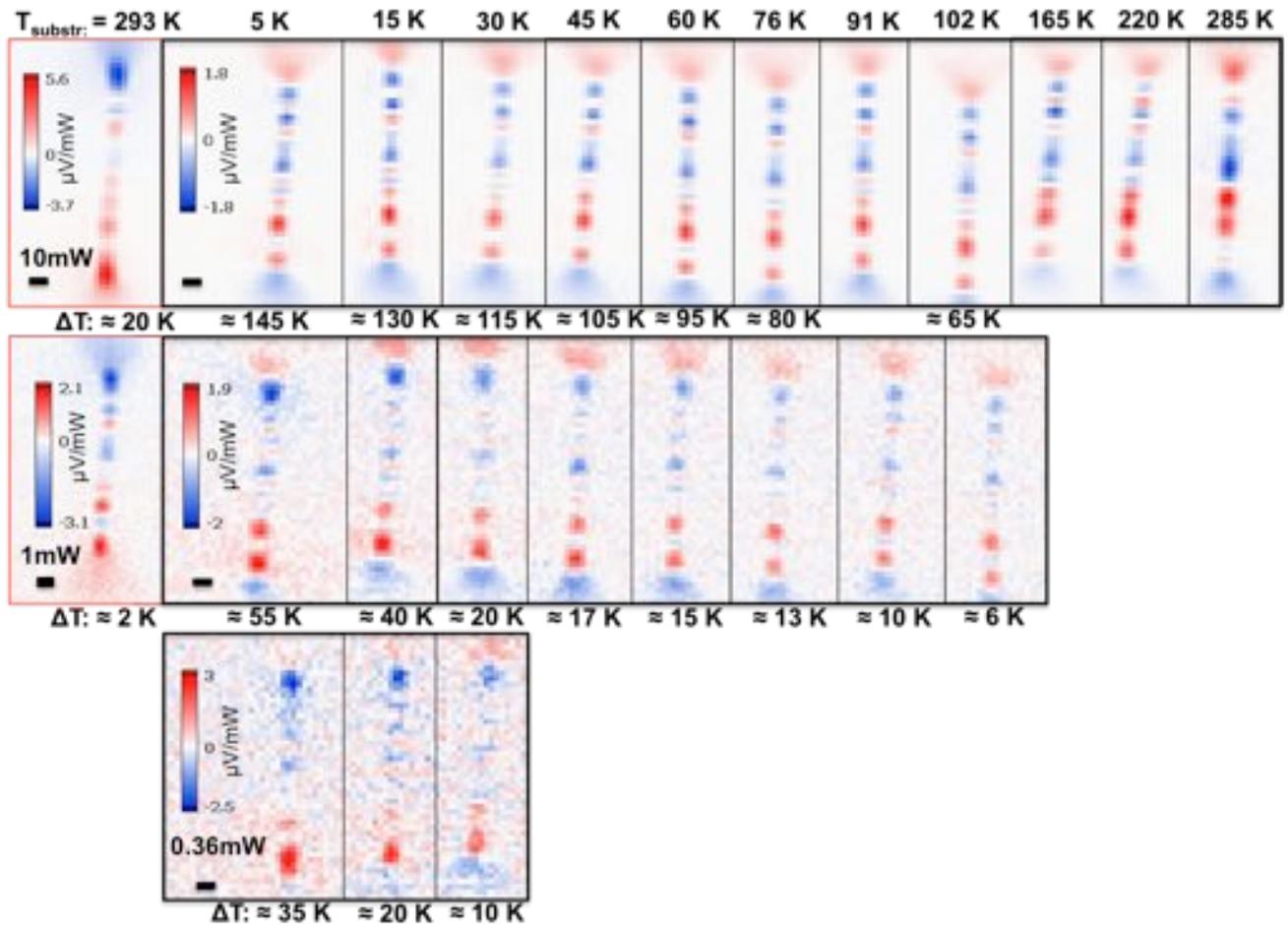

**Fig. S4**. PTE voltage maps for the device from Figure 2 of the main text recorded at different levels of laser power and at different substrate temperatures. Data is organized in rows of voltage maps recorded at the same laser power: 10mW, 1mW, and 0.36 mW. These are arranged by the substrate temperature (indicated at the top) starting from the initial scan at room temperature proceeding to the one at 5 K and through a number of intermediate temperatures to 285 K. The temperature increase in the center of the nanowire was estimated using a bolometric method and is indicated at the bottom of each PTE voltage map. The units in all panels are in µV per mW of laser power on the sample.

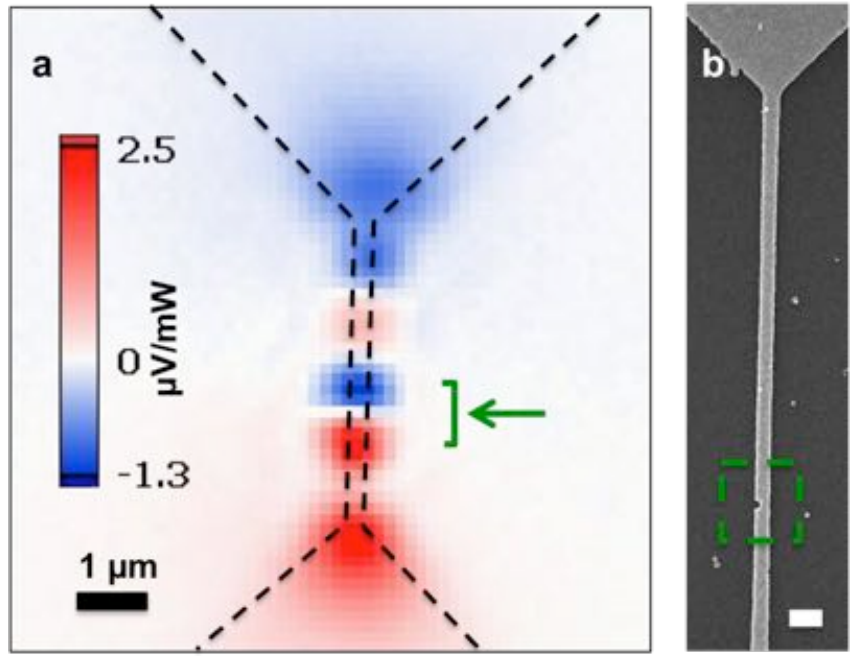

**Fig. S5**. Additional example of the correlation between a small structural defect and the features in the PTE voltage map. (a) Distribution of the PTE voltage recorded at room temperature for the Au device with Ti adhesion layer. Arrow marks the location of the constriction. (b) SEM image of the same device, dashed line highlights position of the defect. Scale bar is 200nm. Note a sign change to positive in the upper half of the nanowire despite the absence of a clearly visible structural defect in the SEM image.

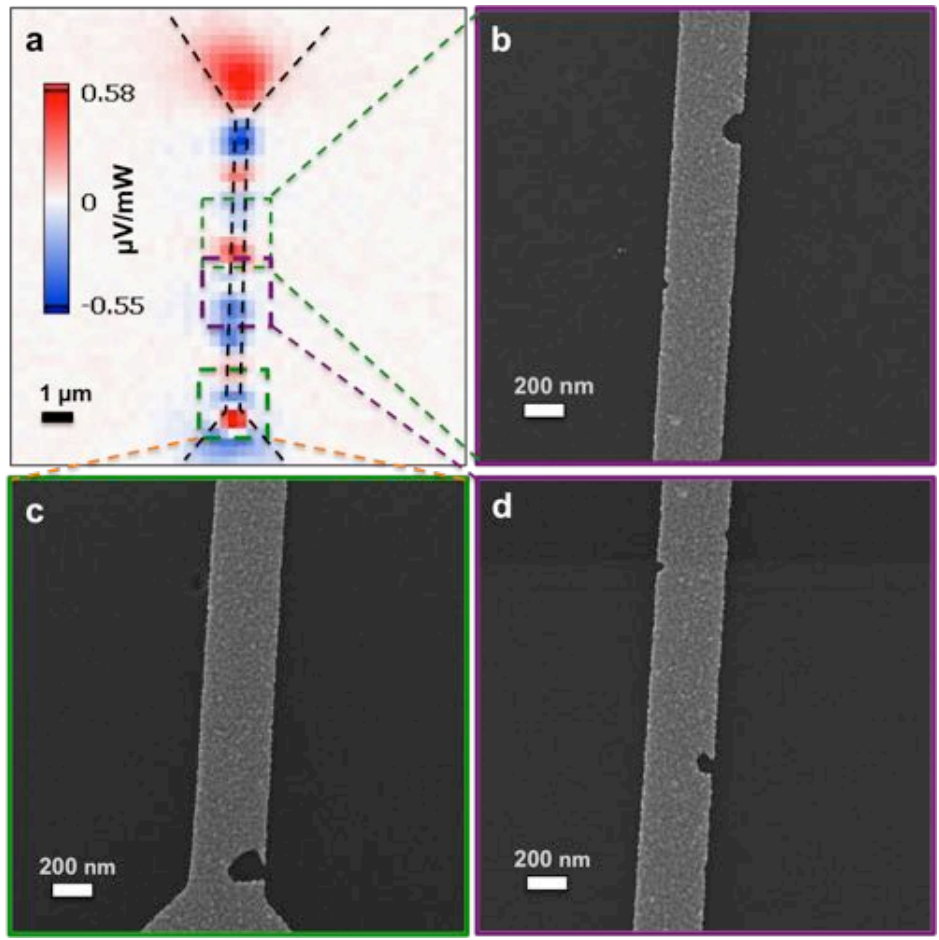

**Fig. S6.** Additional example of the correlation between structural defects and the features in the PTE voltage map for a wide nanowire. (a) Distribution of the PTE voltage recorded at base temperature of 5 K for the Au device with Ti adhesion layer. (b),(c),(d) SEM images of the selected areas of the nanostructure. Dashed lines highlight areas in the PTE voltage map that correspond to the SEM images.

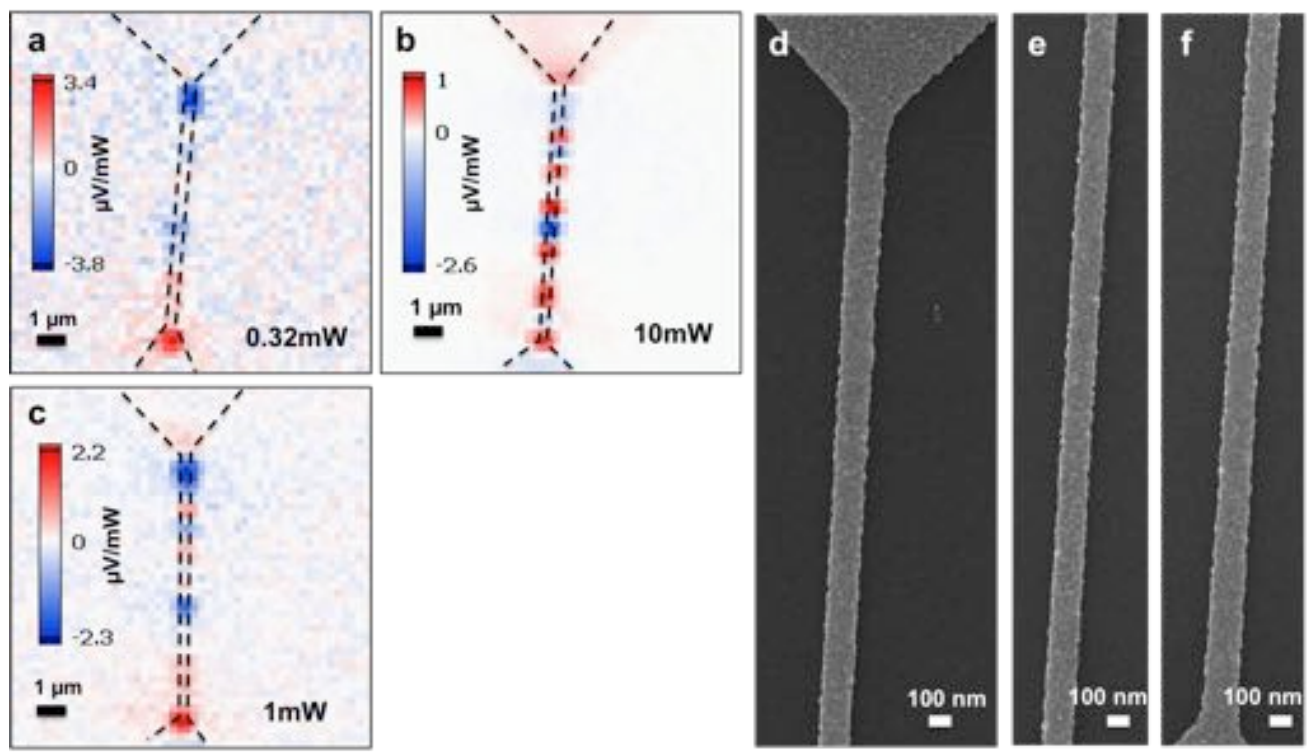

**Fig. S7**. Example of the presence of strong features in the PTE voltage map for a device without clearly defined defects in the nanowire geometry. (a),(b),(c) PTE voltage map of the Au device with Ti adhesion layer measured at substrate temperature of 5 K and different levels of laser power: 0.32, 1, and 10 mW. (d),(e),(f) SEM images of the nanowire that correspond to the top, middle, and bottom of the nanowire. The nanowire width is ~130 nm.

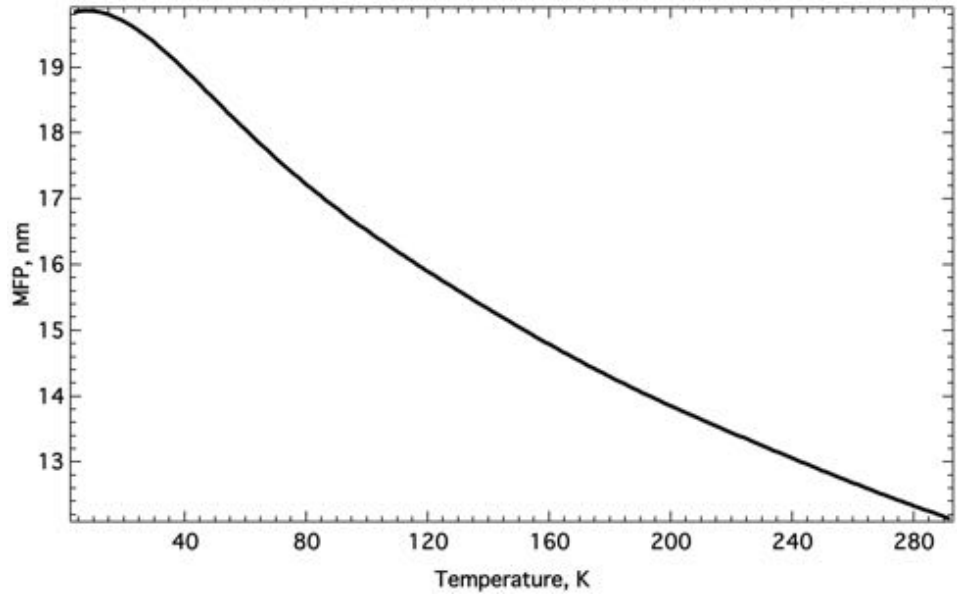

**Fig. S8**. An estimate of the electron mean free path using the Drude model and temperature dependence of the sheet resistance of the gold film evaporated using the same settings as the metallization layer for nanowire fabrication. The sheet resistance was measured using Van der Pauw method. The free electron density of $5.9 \cdot 10^{22} cm^{-3}$, Fermi velocity of $1.4 \cdot 10^6 \, m/s$, and free electron mass were used to make the estimate.

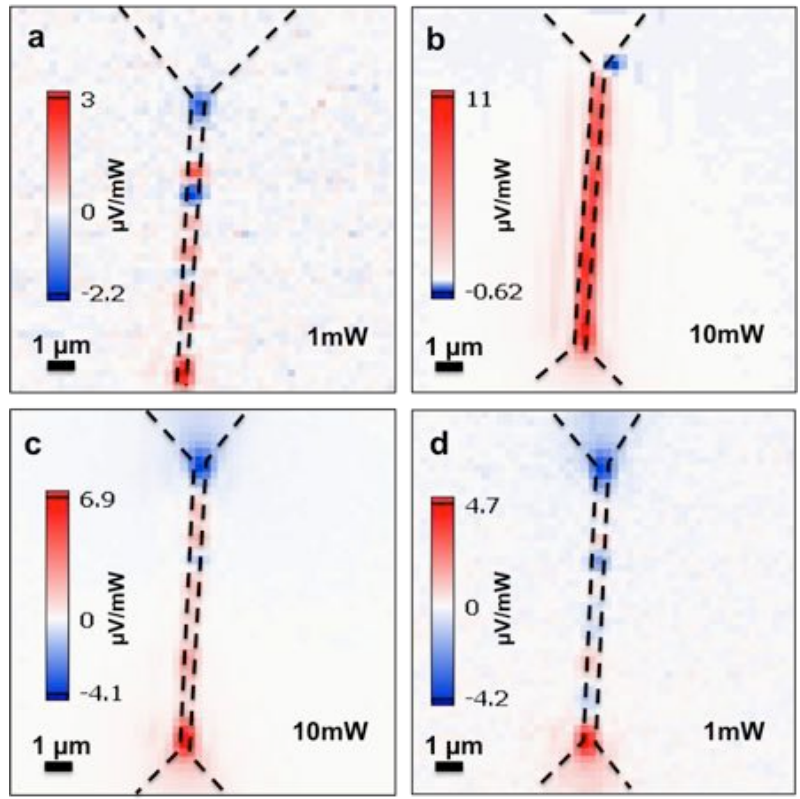

**Fig. S9.** Laser power dependent PTE voltage map for devices after O$_2$ plasma treatment in a 100W barrel cleaner/sterilizer for 3 min. (a) Initial PTE voltage map for the device after the plasma treatment at low laser power of 1 mW. (b) PTE voltage map for the same device at laser power of 10 mW. (c), (d) Sequential PTE voltage maps acquired at laser power of 10 mW and 1 mW. Data acquired at room temperature.

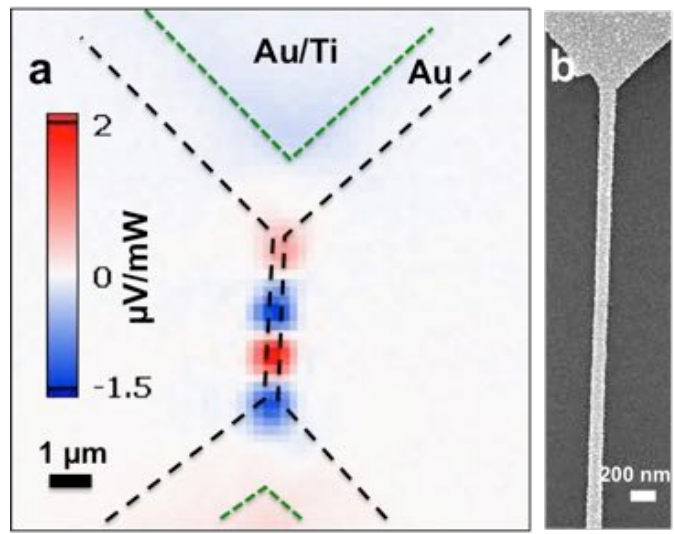

**Fig. S10** Example of the spatial variability of the PTE voltage sign in devices without Ti adhesion layer measured at room temperature. (a) PTE voltage map of the device. Green dashed line denotes location of the larger Au/Ti pads that were used to "anchor" the Au-only film on the SiO₂/Si substrate. (b) SEM image of the device demonstrating lack of clearly visible structural defects.

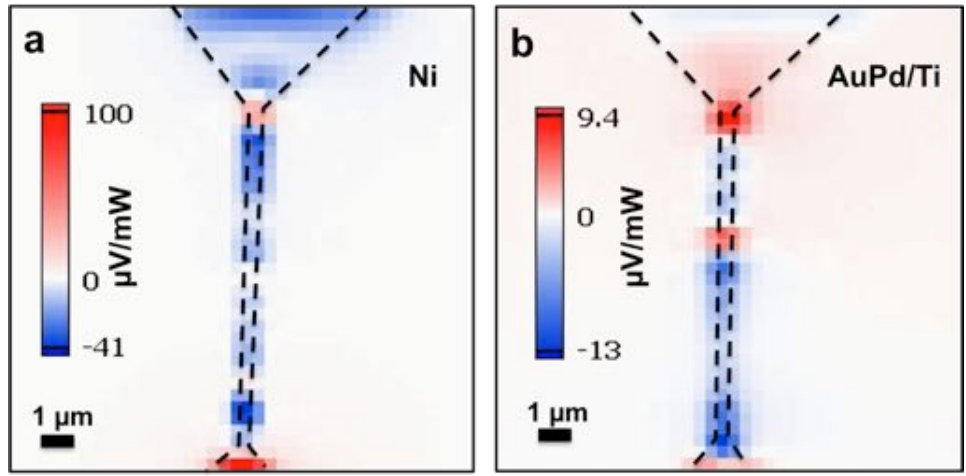

**Fig. S11** Example of the spatial variability of the PTE voltage sign in devices fabricated from Ni metal and AuPd alloy with Ti adhesion layer. The total metallization layer thickness is 15 nm. Data acquired at room temperature.

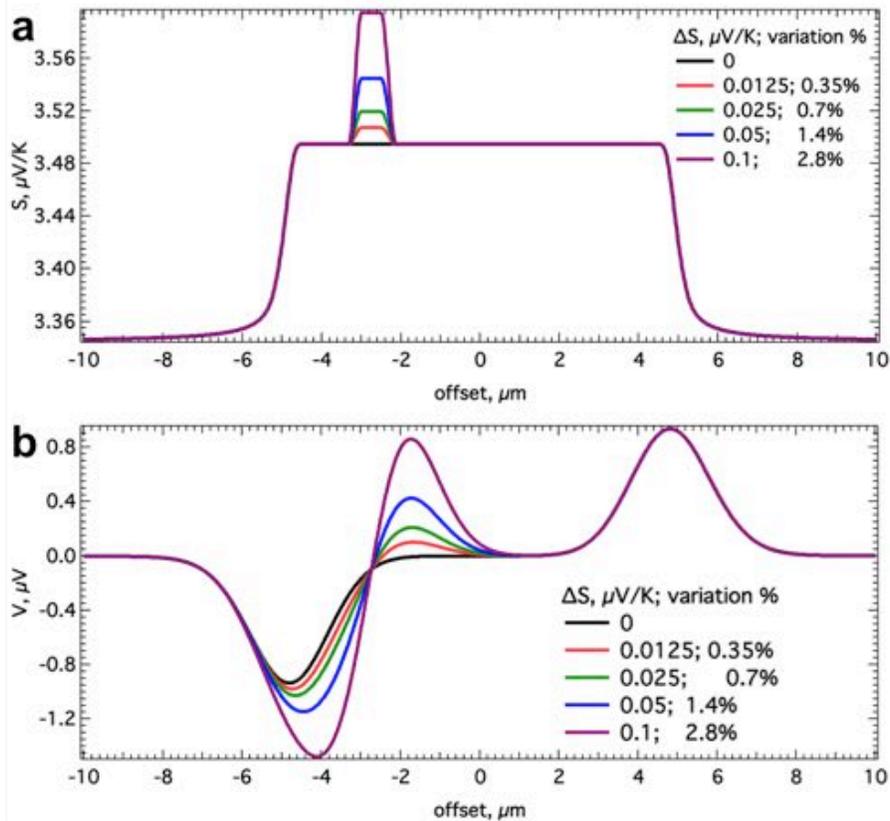

**Fig. S12** Using the model of Fig. S1 to examine the effects of local variations in *S*. We consider a geometry very similar to that shown in Fig. S1, but now a section of the nanowire has a different value of Seebeck coefficient. As the structure is no longer symmetric, the thermoelectric voltage is calculated for both negative and positive values of the heater offset. (a) Local value of S used in the calculation. As a test case for local variation in *S*, we consider a section of the device between -2.5 and 3 μm to have a modified value of Seebeck coefficient, as shown by the *ΔS* value indicated on the graph. (b) The thermoelectric voltage across the device that corresponds the *S* profile demonstrated in (a). For this particular set of parameters, a local variation in *S* of a few percent produces a local voltage inversion (qualitatively similar to that seen in the experiment). When the total amount of heating is considered for the actual experimental conditions and inferred temperature gradients, local *ΔS* of 10-20% is required to reproduce the PTE voltage modulations seen in the experiments.